\documentclass[twocolumn]{NobArticle}
\usepackage{flushend}
\runninghead{Shortened Running Article Title}
\footertext{\textit{WWW MIR} (2025)}

\title{A Multi-Granularity Retrieval Framework for Visually-Rich Documents}

\author{
    Mingjun Xu\textsuperscript{1}, 
    Zehui Wang\textsuperscript{1},  
    Hengxing Cai\textsuperscript{1,2},
    Renxin Zhong\textsuperscript{2}
}

\date{
    \textsuperscript{\textbf{1}}
    DP Technology \\ \textsuperscript{\textbf{2}}
    School of Intelligent Systems Engineering, Sun Yat-Sen University, Shenzhen, China
}

\renewcommand{\maketitlehookd}{%
\begin{abstract}
    \noindent Retrieval-augmented generation (RAG) systems have predominantly focused on text-based retrieval, limiting their effectiveness in handling visually-rich documents that encompass text, images, tables, and charts. To bridge this gap, we propose a unified multi-granularity multimodal retrieval framework tailored for two benchmark tasks: MMDocIR and M2KR. Our approach integrates hierarchical encoding strategies, modality-aware retrieval mechanisms, and vision-language model (VLM)-based candidate filtering to effectively capture and utilize the complex interdependencies between textual and visual modalities. By leveraging off-the-shelf vision-language models and implementing a training-free hybrid retrieval strategy, our framework demonstrates robust performance without the need for task-specific fine-tuning. Experimental evaluations reveal that incorporating layout-aware search and VLM-based candidate verification significantly enhances retrieval accuracy, achieving a top performance score of 65.56. This work underscores the potential of scalable and reproducible solutions in advancing multimodal document retrieval systems.
    
    \medskip

    \small{\textbf{Index Terms:} Multimodal Document Retrieval, Multi-Granularity Modeling, Vision-Language Models, Retrieval-Augmented Generation}
\end{abstract}
}

\begin{document}

\small
\maketitle

\section{Task Description}

Retrieval-augmented generation (RAG) systems are widely adopted to enhance the factual consistency of large language models. However, most existing RAG systems are limited to text-only retrieval, making them ineffective for visually-rich documents that contain a combination of text, images, tables, and charts. To bridge this gap, the Multimodal Document Retrieval (MDR) Challenge\footnote{\url{https://erel-mir.github.io/challenge/mdr-track1/}} introduces two benchmark tasks:

\begin{itemize}
\item \textbf{Task 1: MMDocIR} – Multimodal Page Retrieval, which focuses on retrieving full pages from long documents across various domains such as industry reports and academic papers.
\item \textbf{Task 2: M2KR} – Open-Domain Visual Retrieval, which involves identifying relevant image regions or pages from visually-rich web content like Wikipedia.
\end{itemize}

Both tasks require a unified retrieval system capable of modeling complex interdependencies across textual and visual modalities, handling heterogeneous document layouts, and maintaining fine-grained relevance under long-context settings. Performance is measured using average recall at top-1, top-3, and top-5 rankings~\cite{fu20251sterelmirworkshopefficient}.

Our solution, developed by Team \textbf{LLMHunter}, adopts a training-free hybrid framework that integrates multiple off-the-shelf text and vision-language models in a modular design, enabling robust retrieval without task-specific fine-tuning. By combining diverse retrieval signals—such as semantic text similarity, OCR-enhanced image-text alignment, and layout-aware heuristics—and incorporating a VLM-based candidate filtering module for semantic verification, our method emphasizes robustness, zero-shot generalization, and ease of deployment. Without relying on task-specific training, our approach achieves competitive performance and establishes a strong baseline for multimodal document retrieval.

We detail the overall architecture, model choices, fusion strategies, and experimental insights that guided our design, and discuss how training-free multimodal retrieval can provide a scalable and reproducible solution to real-world multimodal information retrieval challenges.
\section{Related Work}

Multimodal document retrieval has garnered significant attention, leading to the development of models that effectively integrate visual and textual information. ColPali~\cite{faysse2025colpaliefficientdocumentretrieval} introduces a vision-language model that directly embeds document page images into high-quality multi-vector representations, utilizing a late interaction mechanism for efficient retrieval. This approach simplifies the retrieval pipeline by eliminating the need for text extraction and layout analysis, achieving superior performance on the ViDoRe benchmark. Complementing this, GME~\cite{zhang2025gmeimprovinguniversalmultimodal} presents a universal multimodal retrieval framework leveraging multimodal large language models (MLLMs). By constructing a large-scale, high-quality fused-modal training dataset, GME addresses modality imbalance and enables the embedding of diverse inputs—text, images, and their combinations—into a unified vector space, demonstrating state-of-the-art performance on the Universal Multimodal Retrieval Benchmark (UMRB).

To evaluate and benchmark such retrieval systems, MMDocIR~\cite{dong2025mmdocirbenchmarkingmultimodalretrieval} offers a comprehensive dataset comprising long documents with annotations for both page-level and layout-level retrieval tasks. This benchmark facilitates the assessment of systems' abilities to retrieve relevant content across various modalities and granularities. Similarly, M2KR~\cite{lin2024preflmrscalingfinegrainedlateinteraction} provides a suite of datasets designed for training and evaluating general-purpose vision-language retrievers across multiple tasks, including image-to-text and question-to-text retrieval. M2KR serves as a valuable resource for developing and benchmarking models capable of handling diverse multimodal retrieval scenarios.
\section{Methods}
Our overall approach follows a multi-stage retrieval framework based on multi-granularity modeling. Given a text or multimodal query, we first encode both queries and document candidates into a unified embedding space across different levels of granularity, including page-level and region-level representations. Retrieval is initially conducted through similarity-based matching, and fusion strategies are employed to integrate multiple retrieval signals. Finally, a vision-language model (VLM)-based candidate filtering module is used to verify and refine the results for enhanced precision.

Both the M2KR and MMDocIR tasks utilize this multi-granularity design philosophy, with M2KR focusing on fine-grained region retrieval and MMDocIR emphasizing full-page retrieval combined with hierarchical validation. In the following subsections, we present detailed descriptions of our methods for each track.

\subsection{Overall Retrieval Framework}

To address the challenges of multimodal document retrieval, we design a comprehensive retrieval framework that unifies multi-granularity representations and multimodal fusion strategies. The framework is structured into two main components: (1) \textbf{M2KR}, focusing on fine-grained region-level retrieval through integrated visual and textual signals; and (2) \textbf{MMDocIR}, which combines full-page retrieval with detailed regional and OCR-based retrieval stages. Both components share a core architecture of multi-stage matching, similarity-based scoring, and VLM-based candidate filtering to ensure accurate and robust retrieval across diverse document types and query formats. The following sections provide detailed descriptions of these two task-specific solutions.

\subsubsection{M2KR: Multimodal Region Retrieval with Fusion}
In the M2KR task, we propose a fine-grained multimodal retrieval framework that integrates visual and textual signals at the region level. By leveraging multi-granularity representations, our method effectively captures detailed layout and semantic information within document pages to enhance retrieval performance.

\begin{figure}[htbp]
\centering
\includegraphics[width=0.8\columnwidth]{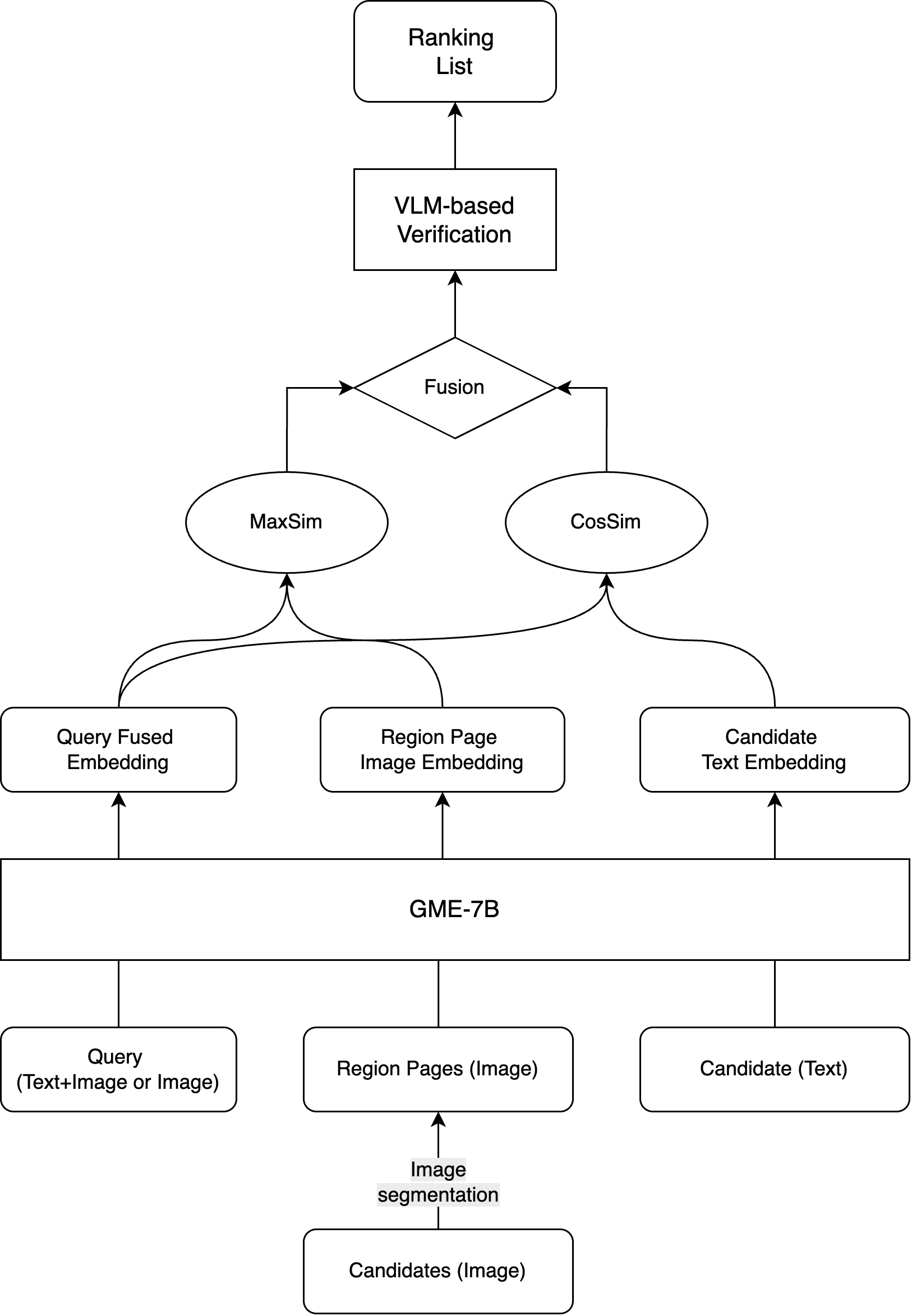}
\caption{Overview of the Multi-Granularity Multimodal Retrieval Framework for M2KR.}
\label{fig:m2kr}
\end{figure}

\textbf{Input:}
The retrieval process starts by preparing multi-granularity inputs. As shown in Figure~\ref{fig:m2kr}, each candidate page is segmented into multiple regions to obtain region-level representations. Simultaneously, text descriptions for the candidates are generated using the \texttt{Qwen2.5-VL} model, providing complementary semantic information. The user query, which can be in text, image, or multimodal format, is also processed into a unified embedding space.

\textbf{Process:}
Three types of matching strategies are applied: pure image retrieval based on region embeddings, multimodal retrieval combining query and region features, and pure text retrieval between query and candidate text descriptions. Cosine similarity (CosSim) is calculated for each modality. The resulting relevance scores from the three matching strategies are then fused to jointly consider visual, multimodal, and textual signals.

\textbf{Output:}
After score fusion, a VLM-based verification module is used to assess the semantic alignment between the query and top candidates. This filtering stage promotes high-confidence matches while discarding false positives, ensuring the final results are both precise and semantically faithful to the query.

\subsubsection{MMDocIR: Full-Page Retrieval with Multistage Validation}
In the MMDocIR task, we design a hierarchical retrieval framework that combines full-page and regional information. By integrating global page-level retrieval with fine-grained regional matching and OCR text retrieval, our method achieves robust document understanding and significantly improves retrieval accuracy.

\begin{figure}[htbp]
\centering
\includegraphics[width=1.0\columnwidth]{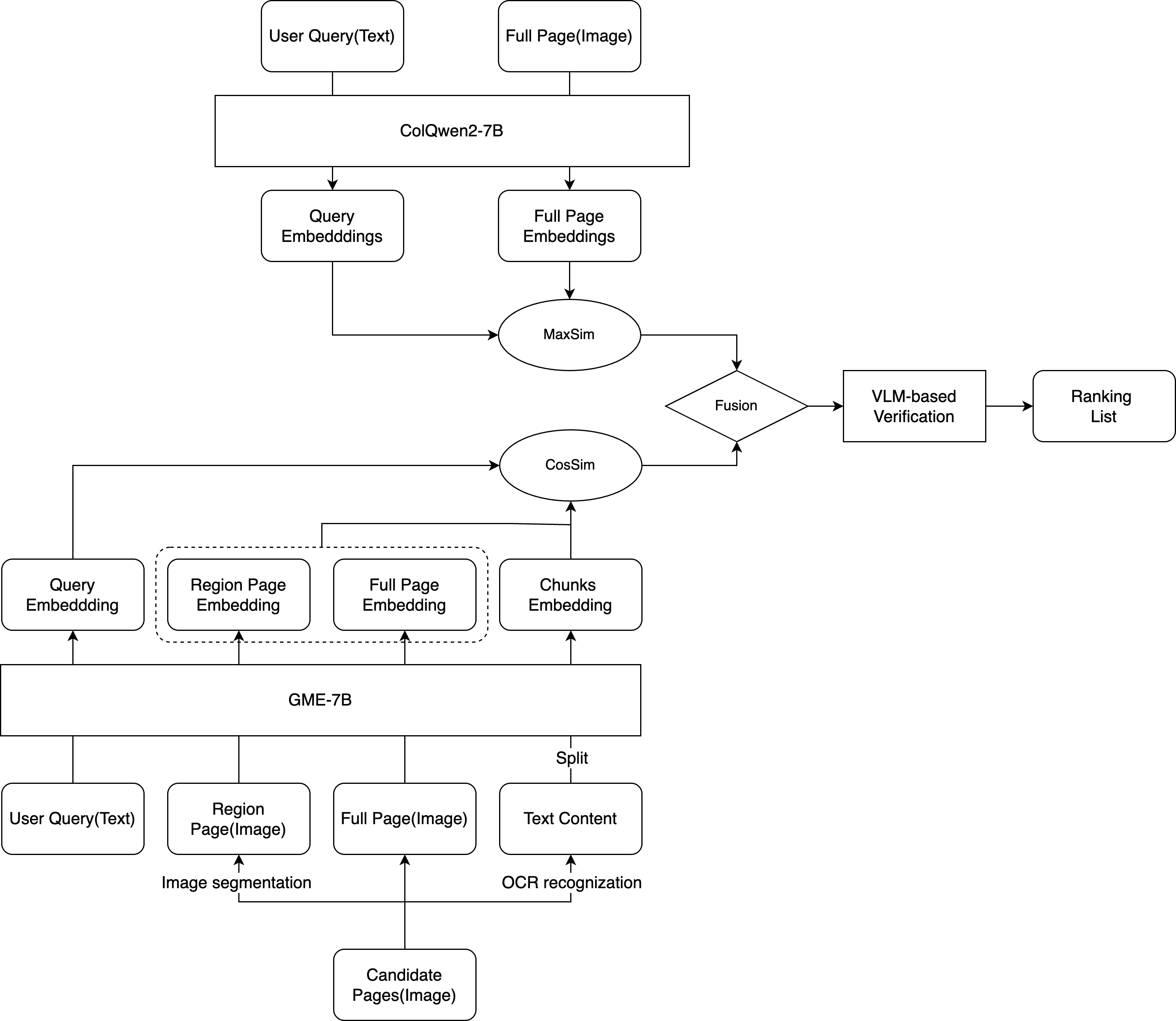}
\caption{Overview of the MMDocIR retrieval framework combining full-page and regional information.}
\label{fig:mmdocir}
\end{figure}

\textbf{Input:}
The retrieval process begins with a user query provided in text form. As shown in Figure~\ref{fig:mmdocir}, candidate document pages are prepared with their full-page images, segmented region images, and OCR-recognized texts to enable multi-granularity retrieval.

\textbf{Process:}
First, ColQwen2-7B encodes the user query and full-page image candidates to perform initial full-page retrieval. Simultaneously, we apply GME-7B for two additional retrieval paths: (1) text-to-OCR-text retrieval, where the query is matched against the OCR texts extracted from the pages; and (2) text-to-region-image retrieval, where the query is matched against segmented regional images. Cosine similarity (CosSim) is calculated for each retrieval path. The relevance scores obtained from full-page retrieval, OCR-text retrieval, and regional-image retrieval are fused to combine global, textual, and regional information.

\textbf{Output:}
After fusion, a VLM-based verification module—powered by the Qwen2.5-VL model—is applied to semantically validate the top-ranked candidates. Rather than merely reordering, this module conducts fine-grained cross-modal verification to confirm the alignment between the query intent and candidate content, ensuring the final results maintain high precision and reliability.

\subsection{Fusion and Verification Mechanism}

To maximize retrieval accuracy, we employ a combined strategy of multi-source score fusion and vision-language-based verification. This ensures that both diverse retrieval paths and final semantic alignment are fully leveraged.

\subsubsection{Multi-Source Score Fusion}

We implement a two-layer score fusion process. First, in the modality score integration stage, the GME model generates three types of relevance scores for each candidate: text-to-OCR-text similarity, text-to-region-image similarity, and multimodal similarity between the query and region features. These scores capture different aspects of document relevance and are directly used for initial ranking, providing a rich multi-granularity matching signal.

Second, to combine the outputs from both the GME and ColQwen retrieval paths, we apply Reciprocal Rank Fusion (RRF). This method merges ranked lists by assigning higher weight to top-ranked results from each retrieval path, effectively balancing global and fine-grained retrieval signals. The fused ranking ensures that strong candidates from different models are jointly considered, enhancing retrieval robustness.

\subsubsection{VLM-Based Candidate Verification}

After the fusion step, we apply a vision-language-based verification process using the Qwen2.5-VL model. This module performs semantic verification by predicting whether each candidate is a true match (\texttt{Yes}) or not (\texttt{No}). Verified positive candidates are prioritized in the final output, ensuring that only results with strong semantic alignment are retained. This verification step adds an essential layer of cross-modal understanding, refining the retrieval results beyond traditional similarity-based methods.
\section{Results and Analysis}

\begin{figure}[htbp]
\centering
\includegraphics[width=0.95\columnwidth]{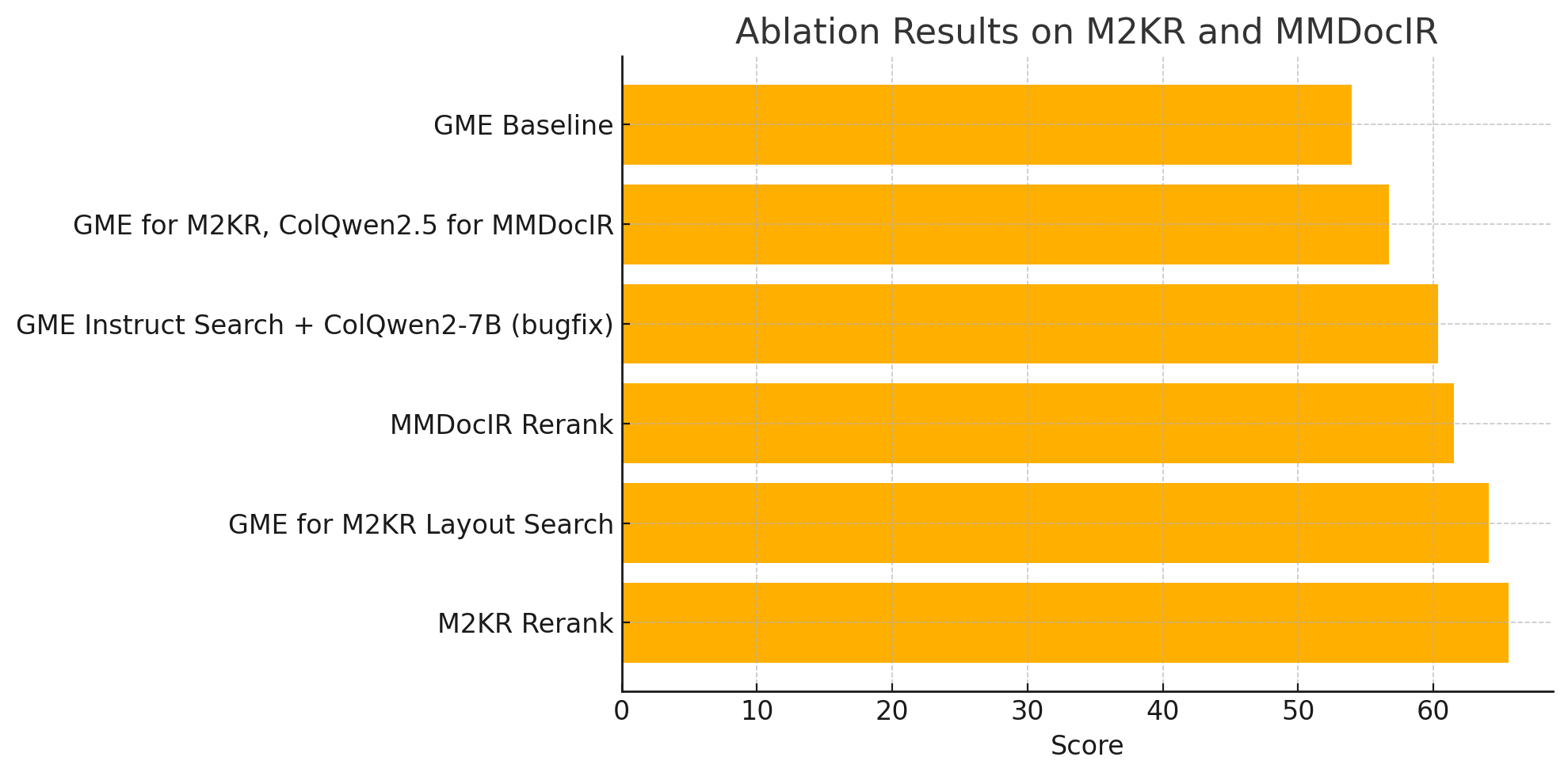}
\caption{Performance trend across progressive improvements.}
\label{fig:ablation}
\end{figure}

In this section, we present the experimental results of different retrieval configurations on the M2KR and MMDocIR tasks. We evaluate the progressive improvements brought by integrating advanced modules such as ColQwen2.5, layout-aware search, and reranking strategies. Table~\ref{tab:ablation} and Figure~\ref{fig:ablation} summarize the performance of each variant in terms of retrieval score.

\begin{table}[htbp]
\centering
\caption{Ablation results on M2KR and MMDocIR tasks.}
\label{tab:ablation}
\begin{tabular}{ll}
\toprule
\textbf{Method} & \textbf{Score} \\
\midrule
GME Baseline & 53.9613 \\
GME for M2KR, ColQwen2.5 for MMDocIR & 56.6967 \\
GME Instruct Search + ColQwen2-7B (bugfix) & 60.3278 \\
MMDocIR Rerank & 61.5289 \\
GME for M2KR Layout Search & 64.0902 \\
M2KR Rerank & \textbf{65.5588} \\
\bottomrule
\end{tabular}
\end{table}

The results show a clear upward trend as we incorporate task-specific enhancements. Applying ColQwen2.5 and fixing multimodal bugs led to a substantial boost in MMDocIR, while introducing layout-aware search and reranking significantly benefited M2KR. Our final configuration, which integrates reranking for both tasks, achieves the highest retrieval score of 65.56.


\section{Conclusion}

In this work, we propose a unified multi-granularity multimodal retrieval framework and apply it to two document-level tasks: M2KR and MMDocIR. By leveraging hierarchical encoding, modality-aware retrieval strategies, and vision-language-based candidate verification, our approach progressively enhances retrieval accuracy. Experimental results show a clear performance gain from the integration of layout-aware search and VLM-based verification modules, with the final system achieving a score of 65.56. These results demonstrate the effectiveness and scalability of our framework in complex multimodal document retrieval scenarios.

\printbibliography

\end{document}